\documentclass[
    aps,
    floatfix,
    prb,
    twoside,
    twocolumn,
    10pt,
    showpacs,
    citeautoscript,
    superscriptaddress,
]{revtex4-1}

\usepackage[version=4]{mhchem}
\usepackage{amsmath}
\usepackage{amssymb}
\usepackage{booktabs}
\usepackage{glossaries}
\usepackage{graphicx}
\usepackage{tabularx}
\usepackage{siunitx}
\usepackage[dvipsnames]{xcolor}

\usepackage[colorlinks=true]{hyperref}
\hypersetup{
    pdffitwindow=false,
    pdfstartview={FitH},
    pdfnewwindow=true,
    colorlinks=true,
    linkcolor=NavyBlue,
    citecolor=NavyBlue,
    filecolor=NavyBlue,
    urlcolor=NavyBlue,
}

\hypersetup{pdfauthor={C. Linderälv, J. Hagel, S. Brem, E. Malic, and P. Erhart}}
\hypersetup{pdftitle={The moiré potential in twisted transition metal dichalcogenide bilayers}}

\setacronymstyle{long-short}
\newacronym{cbm}{CBM}{conduction band minimum}
\newacronym{dft}{DFT}{density functional theory}
\newacronym{lda}{LDA}{local density approximation}
\newacronym{si}{SI}{Supplementary Information}
\newacronym{tmd}{TMD}{transition metal dichalcogenide}
\newacronym{vbm}{VBM}{valence band maximum}
\newacronym{xc}{XC}{exchange-correlation}

\newcommand{\addchalmers}{Department of Physics, Chalmers University of Technology, SE-41296, Gothenburg, Sweden}
\newcommand{\addmarburg}{Department of Physics, Philipps University of Marburg, 35037 Marburg, Germany}

\begin{document}

\title{The moir\'e potential in twisted transition metal dichalcogenide bilayers}

\author{Christopher Linder\"alv}
\affiliation{\addchalmers}
\author{Joakim Hagel}
\affiliation{\addchalmers}
\author{Samuel Brem}
\affiliation{\addmarburg}
\author{Ermin Malic}
\affiliation{\addmarburg}
\affiliation{\addchalmers}
\author{Paul Erhart}
\email{erhart@chalmers.se}
\affiliation{\addchalmers}

\begin{abstract}
Moir\'e superlattices serve as a playground for emerging phenomena, such as localization of band states, superconductivity, and localization of excitons.
These superlattices are large and are often modeled in the zero angle limit, which obscures the effect of finite twist angles.
Here, by means of first-principles calculations we quantify the twist-angle dependence of the moir\'e potential in the \ce{MoS2} homobilayer and identify the contributions from the constituent elements of the moir\'e potential.
Furthermore, by considering the zero-angle limit configurations, we show that the moir\'e potential is rather homogeneous across the \glspl{tmd} and briefly discuss the separate effects of potential shifts and hybridization on the bilayer hybrid excitons.
We find that the moir\'e potential in \glspl{tmd} exhibits both an electrostatic component and a hybridization component, which are intertwined and have different relative strengths in different parts of the Brillouin zone.
The electrostatic component of the moir\'e potential is a varying dipole field, which has a strong twist angle dependence.
In some cases, the hybridization component can be interpreted as a tunneling rate but the interpretation is not generally applicable over the full Brillouin zone.
\end{abstract}

\maketitle

\section{Introduction}

Hexagonal bilayer \acrfullpl{tmd} with a rotational misorientation between the monolayer sheets can exhibit an interference pattern in the atomic positions known as a moir\'e pattern.
The latter can alter the electronic structure of the twisted \gls{tmd} bilayer and lead to, e.g., localized electronic \cite{NaiJai18} and excitonic states \cite{BreLinErh20}.
Since the moir\'e unit cell at low twist angles (for a lattice constant matched bilayer) becomes very large, such structures are very demanding to simulate with first-principles calculations. 

\Gls{dft}-based first-principles calculations of the \emph{ground state} of moir\'e structures have revealed localized electronic states \cite{NaiJai18}.
It is, however, computationally much more difficult to access electronic \emph{excitations} within such a framework.
Hence, tight-binding models \cite{YuLiuTan17} remain the most feasible route to access electronic properties of twisted bilayers at low angles.

\begin{figure*}
    \centering
    \includegraphics[width=\linewidth]{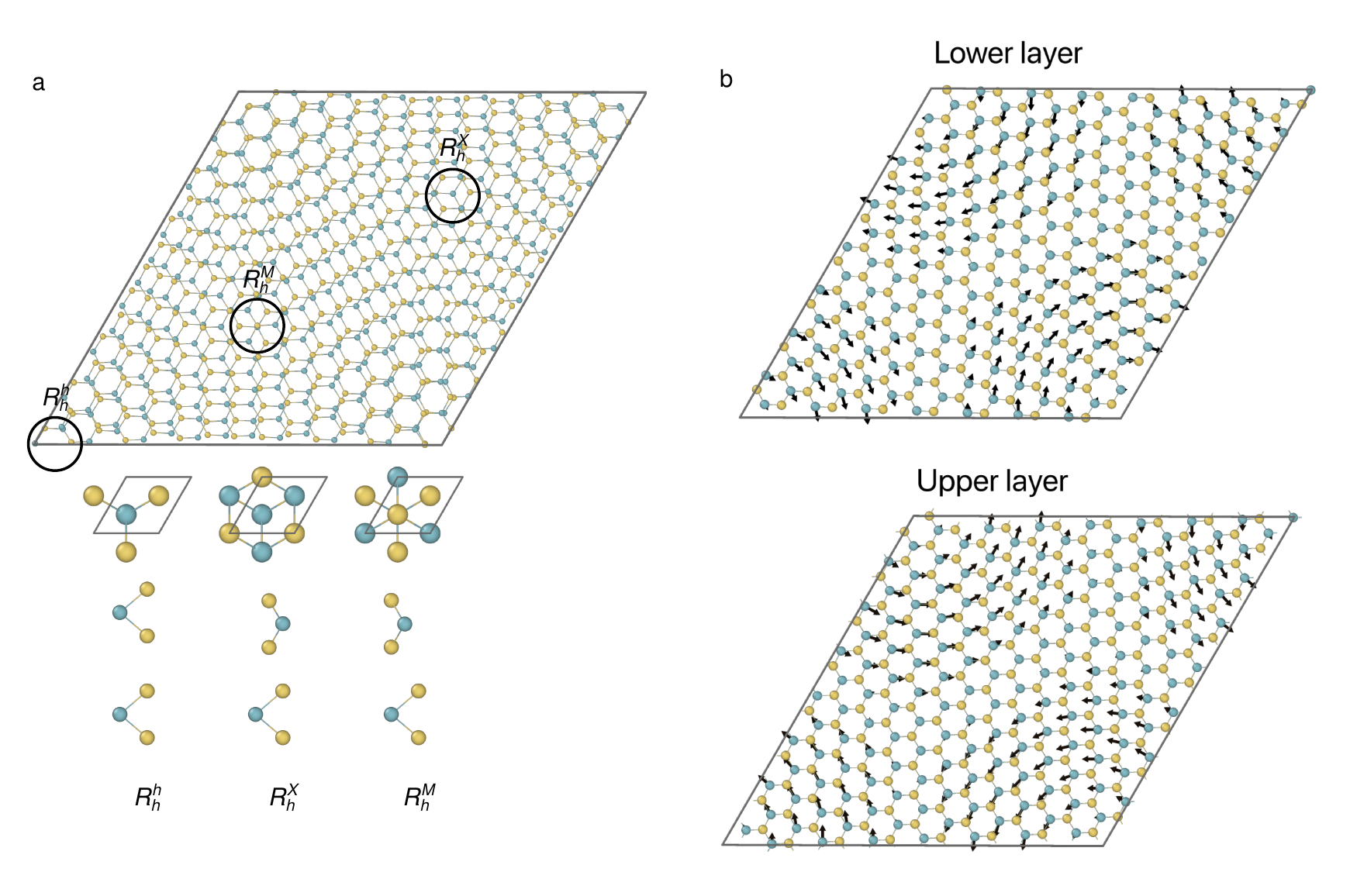}
    \caption{
        (a) The explicit \ce{MoS2} homobilayer moir\'e structure at a twist angle of $\theta=2.9^{\circ}$ along with the limiting ($\theta\rightarrow 0^{\circ}$) local registry at three specific points.
        Blue (yellow) atoms indicate atomic species Mo (S).
        (b) Lateral relaxation for the \ce{MoS2}/\ce{MoS2} homobilayer twisted 4.41$^{\circ}$.
        The arrows are amplified with a factor of 30.
    }
    \label{fig:local-registry}
\end{figure*}

The twisting induces a moir\'e pattern with spatially varying local stacking orders (\autoref{fig:local-registry}a).
There are three locations within the moir\'e unit cell where the local environment exhibits high symmetry (threefold rotational symmetry), which  can be modeled by stacking the atoms accordingly in the primitive bilayer cell (\autoref{fig:local-registry}a).
This method for analyzing twist-induced phenomena in \glspl{tmd} has been used in, e.g., Refs.~\citenum{BreLinErh20, YuLiuTan17, RuiFal19, LuLiYan19}.
The caveat of this method is that it is strictly valid only at very low twist angles since for shorter moir\'e periods the atomic arrangement changes rapidly away from the high symmetry points.
Therefore, it is of importance to quantify the twist angle dependence of the induced potential. 

Furthermore, in order to construct more accurate tight-binding models the electrostatic part of the moir\'e potential needs to be disentangled from the tunneling contribution, which calls for a deeper investigation of the electronic structure of the limiting configurations. 
This includes a more careful study of the actual origin and nature of the moir\'e potential, and recently several studies emerged targeting the fundamental properties of the moir\'e potential.
Specifically, the moir\'e potential in \ce{MoSe2}/\ce{WSe2} heterobilayers has been shown to be on the order of 150 to \SI{300}{\milli\electronvolt} and dominated by planar strain \cite{ShaHalWu21}, and the band gap in \ce{MoS2}/\ce{MoTe2} has been observed to be strongly twist-angle dependent \cite{GenWanLin21}.
A comprehensive description that connects the aforedescribed observations and establishes general trends that hold across different combinations of \glspl{tmd} is, however, still needed.

Here, we therefore assess the moir\'e potential via first-principles calculations both on explicit moir\'e configurations and limiting configurations based on the primitive unit cell (\autoref{fig:local-registry}).
We aim to provide a unifying perspective of the moir\'e potential in \glspl{tmd}, in particular of the origin of the potential and the similarity of the potential in different bilayer systems.
Finally, we disentangle the electrostatic part of the moir\'e potential and determine the hybridization contribution at high symmetry points in the moir\'e superlattice. 

\section{Methodology}

In order to construct the explicit moir\'e superlattices we have used the method outlined in Ref.~\citenum{LopPerCas07}.
The commensurate hexagonal moir\'e superlattices are found at angles \cite{LopPerCas07}
\begin{equation}
    \theta_k = \arccos \left(\frac{3k^2+3k+0.5}{3k^2+3k+1}\right),
\end{equation}
where $k$ is a non-negative integer.
In this work, we consider values up to $k=7$, corresponding to an angle of $4.41^{\circ}$.
The variations in the local atomic environment give rise to an electrostatic potential that varies in the plane of the monolayer sheets, which we denote as the \emph{intralayer} moir\'e potential.
By definition, the difference of the \emph{intralayer} moir\'e potentials in the two adjacent layers is the \emph{interlayer} moir\'e potential \cite{GenWanLin21}.

We consider two different measures of the electrostatic potential induced by twisting:
(1)~The (average) electrostatic core potential is a measure of the local electrostatic environment of the relevant electronic states, which in \glspl{tmd} are primarily composed of $d$ orbitals of the transition metal \cite{KorBurGmi15}.
(2)~The second measure, which is closely related to the first is the induced dipole field \cite{TonCheXia20}.
It is strictly applicable only to the limiting configurations, since in explicit moir\'e superlattices, the field exhibits higher moments.
The charge density difference upon bilayer formation is 
\begin{align}
    \delta n(\boldsymbol{r}) = n(\boldsymbol{r})-\sum_in_i(\boldsymbol{r}),
    \label{eq:density-difference}
\end{align}
where $n(\boldsymbol{r})$ is the charge density of the bilayer structure and $n_i(\boldsymbol{r})$ denotes the charge densities of the constituent monolayers.
The charge density difference $\delta n$ gives rise to a potential $\delta V$.
In order to accurately determine the potential $\delta V$ and the induced vacuum level difference $D$
\begin{equation}\label{eq:D}
    D=\delta V(-\infty) -\delta V(\infty),
\end{equation}
dipole corrections are added when solving the Poisson equation for the electron density.
 
In order to estimate how $\delta V$ affects different single particle states we compute the following matrix elements for the monolayer states
In order to estimate how $\delta V$ affects different single particle states we compute the following matrix elements for the monolayer states
\begin{align}
     M_{n\boldsymbol{k}}^{L} = \langle \psi^L_{n\boldsymbol{k}}|\delta V|\psi^L_{n\boldsymbol{k}}\rangle,
     \label{eq:m}
\end{align}
where $L$ is the monolayer index and $n$ is either $v$ (valence band) or $c$ (conduction band).
In the spirit of perturbation theory, $M_{n\boldsymbol{k}}^{L}$ is the first-order shift of the state $n\boldsymbol{k}$ in monolayer $L$ due to the induced electrostatic potential.
The (average) hybridization contribution to the level shift upon bilayer formation is then computed as 
\begin{align}
    \Delta_{n\boldsymbol{k}} = \frac{1}{2}\Big[
    &\varepsilon_{n\boldsymbol{k}}^{\text{bilayer}, +}-\varepsilon_{n\boldsymbol{k}}^{\text{bilayer}, -} \nonumber\\
    &-\left(M^+_{n\boldsymbol{k}} -M^-_{n\boldsymbol{k}} + \Delta \varepsilon^{\text{mono}}_{n\boldsymbol{k}}\right)\Big],
    \label{eq:deltank}
\end{align}
where $\varepsilon_{n\boldsymbol{k}}^{\text{bilayer, +}}$ is the energy of the bilayer state that shifts towards larger energies, $M^+_{n\boldsymbol{k}}$ is the matrix element in the layer which experiences a positive potential, and $\Delta \varepsilon^{\text{mono}}_{n\boldsymbol{k}}$ is the energy level difference between the monolayers (which is zero for homobilayers).
We will also make use of the (average) interlayer tunneling rate
\begin{align}
T_{n\boldsymbol{k}}^2 = \frac{1}{2}\bigg[
& \left(\varepsilon_{n\boldsymbol{k}}^{\text{bilayer},+}
- \varepsilon_{n\boldsymbol{k}}^{\text{bilayer}, -}\right)^2 \nonumber
\\
&-\left(M^+_{n\boldsymbol{k}} -M^-_{n\boldsymbol{k}}+\Delta \varepsilon^{\text{mono}}_{n\boldsymbol{k}}\right)^2\bigg],
\label{eq:t}
\end{align}
which may be interpreted as the off-diagonal perturbation of a two-level system.
These tunneling strengths may be, e.g., included in a tight-binding model for the exciton energies of twisted structures \cite{BreLinErh20}.

\begin{figure}
    \centering
    \includegraphics{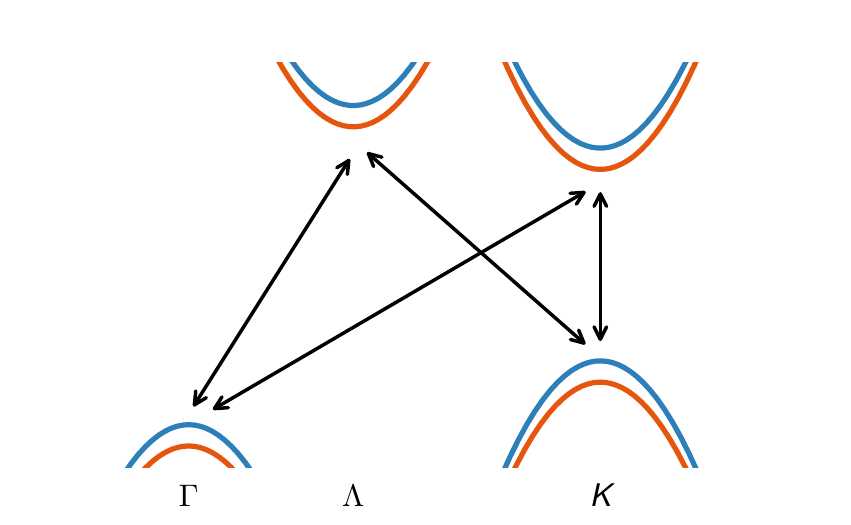}
    \caption{
        Schematic illustration of the possible excitonic transitions considered here with a band arrangement that corresponds to the case of non-interacting monolayers with type II band alignment.
        The $K-K$ exciton is a direct exciton, whereas the other considered excitons ($K-\Lambda$, $\Gamma-\Lambda$, $\Gamma-K$) are indirect in momentum.
    }
    \label{fig:schematic-exciton}
\end{figure}

For the purpose of this paper, we consider a simpler case of untwisted structures to disentangle the effect of electrostatic effects and hybridization \cite{HagBreLin21}.
To this end, we use the exciton density matrix formalism \cite{IvaHau93, SelBerRic18, KatSelCar18} and calculate the exciton energies for untwisted van-der-Waals heterostructures by first formulating a Hamiltonian in second quantization for the interaction free electronic energies
\begin{equation}\label{eq:hamiltonian}
    H = \sum_{\alpha l \boldsymbol{k}} \tilde{E}_{l\boldsymbol{k}}^{\alpha s}a_{\alpha l \boldsymbol{k}}^{\dagger}a_{\alpha l \boldsymbol{k}}^{\phantom{\dagger}} + \sum_{\alpha \boldsymbol{k} l\neq l'}T_{ll'}^{\alpha s}a_{\alpha l \boldsymbol{k}}^{\dagger}a_{\alpha l \boldsymbol{k}}^{\phantom{\dagger}},
\end{equation}
where $l$, $l'$ are layer indices, $\alpha = (\lambda, \xi)$ is a compound index with $\xi$ denoting the valley and $\lambda = (c,v)$ the conduction and the valence band respectively.
Here $a^{\dagger}$ ($a$) denotes the electronic creation (annihilation) operator.
The possible excitonic transitions in this way are illustrated in \autoref{fig:schematic-exciton}.
Here, $\boldsymbol{k}$ is the electronic wave vector and $s$ is the stacking index. 
$\tilde{E}_{l\mathbf{k}}^{\alpha s}$ are the electronic energies as well as the electrostatically induced alignment shifts. 
$T_{ll'}^{\alpha s}$ is the tunneling matrix element, which is modeled following a tight-binding approach \cite{BreLimGil20, WanWanYao17}.
Consequently, the tunneling matrix element are directly proportional to the tunneling strength for those stackings, for which tunneling is allowed by $C_3$ symmetry.
By following the method laid out in Refs.~\citenum{BreLimGil20} and \citenum{BreLinErh20}, the Hamiltonian is transformed into an exciton basis, thus taking into account the binding energies.
The final hybrid exciton energies are calculated by first transforming into a hybrid basis $Y_{\xi\eta\bm{Q}}^{\dagger}=\sum_L\mathcal{C}^{\xi\eta}_{L}(\bm{Q})^*X_{\xi L \bm{Q}}^{\dagger}$, which consists of a linear combination of the intra and interlayer exciton contributions $\mathcal{C}^{\xi\eta}_{L}(\bm{Q})$ to the final hybrid state.
Here, $L$ is a compound layer index, $\bm{Q}$ is the center-of-mass momentum, $\eta$ the new band index and $X^{\dagger}$(X) are the exciton creation(annihilation) operators.
The eigenvalue problem that arises in order to diagonalize this new hybrid exciton Hamiltonian yields the final exciton energies.

\Gls{dft} calculations were performed using the projector augmented wave method\cite{Blo94} while atomic configurations were prepared and analyzed using \textsc{ase} \cite{HjoMorBlo17}.
All structural relaxations were carried out using \textsc{vasp} \cite{KreHaf93, KreFur96} and the vdW-DF-cx method \cite{DioRydSch04, BerHyl14}.
The monolayer lattice parameters were obtained using a plane wave cutoff of \SI{340}{\electronvolt} and a $\boldsymbol{k}$-point mesh with a density of \SI{0.3}{\per\angstrom}.
The atomic positions for the limiting bilayer structures (\autoref{fig:local-registry}) were relaxed using the same functional but with a cutoff energy of \SI{500}{\electronvolt} and a $18\times18\times 1$ $\boldsymbol{k}$-point mesh until the maximum force acting on any atom was less than \SI{2.5}{\milli\electronvolt\per\angstrom}.
The explicit moir\'e superlattices with up to 1,014 atoms were relaxed using a plane wave cutoff of \SI{340}{\electronvolt} until the maximal force acting on any ion was less than \SI{7.5}{\milli\electronvolt\per\angstrom}.
The cell vector in the out-of-plane direction was \SI{70}{\angstrom} for these configurations to ensure decoupling across periodic boundary conditions.

The analysis of the electronic densities and potentials was carried out using \textsc{gpaw} \cite{MorHanJac05, EnkRosMor10} and the \gls{lda} \cite{PerZun81}.
The electron densities of the limiting bilayer configurations were computed using a grid expansion with \SI{0.15}{\angstrom} spacing and a $18\times18\times 1$ $\boldsymbol{k}$-point mesh.
Here, a compensating dipole layer was included in the Poisson solver.

\section{Results}
The \glspl{tmd} with MX$_2$ (M=Mo,W;X=S,Se) constitute a class of materials with similar behavior and here we therefore only carry out computations for large explicit moir\'e superlattices for the MoS$_2$/MoS$_2$ homobilayer.
Computations and analysis of the limiting bilayer stackings have, however, been performed for all homobilayers and lattice constant matched heterobilayers.

\subsection{Structural properties}

First we report the computations of the interlayer distance, which is of fundamental importance since the wave function overlap of the constituent monolayer states is sensitively dependent on the interlayer distance.
Therefore, both polarization and hybridization magnitude depend on the interlayer distance.
In the zero-angle limit configurations, the interlayer distance for \ce{MoS2}/\ce{MoS2} varies between \SI{6.86}{\angstrom} in $R_h^h$ and \SI{6.10}{\angstrom} in $R^X_h$ and $R^M_h$ in decent agreement with other theoretical estimations \cite{LiuXhaCao14}.

\begin{figure*}
    \centering
    \includegraphics[width=\linewidth]{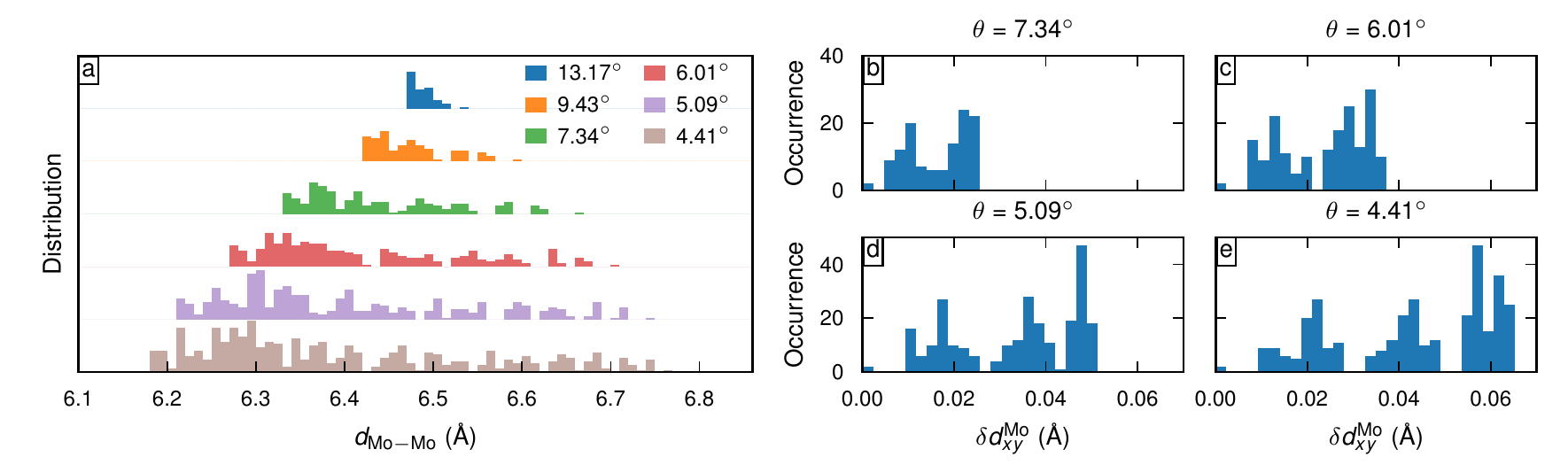}
    \caption{
        (a) Distribution of interlayer distances in \ce{MoS2}/\ce{MoS2} homobilayer as a function of twist angle.
        The lower and upper limits of the x-axis correspond to the layer spacing in $R^h_h$ and $R^M_h$/$R^X_h$, respectively.
        (b--e) The distribution of lateral relaxation for the four lowest twist angles.
    }
    \label{fig:structural-properties}
\end{figure*}

In moir\'e superlattices the interlayer distance varies with the lateral position.
In order to quantify the interlayer distance for the explicit moir\'e structures, we therefore consider the distribution of distances (\autoref{fig:structural-properties}a), which we obtain by centering a cylinder of radius \SI{3}{\angstrom} on a transition metal atom in the upper layer and then taking the interlayer distance as the vertical distance between transition metal atoms found within the same cylinder.

At the largest considered twist angle (13$^{\circ}$) the interlayer distance is rather homogeneous with a value roughly equal to the mean of the interlayer distance in $R_h^h$ and $R^X_h$.
This is not surprising since the interlayer interaction is very weak due to its van-der-Waals nature while the deformation of the monolayer sheets requires bending the much stronger covalent bonds.
As the twist angle decreases the distribution widens.
The extrema occur at $R_h^h$ and $R_h^M$ ($R^X_h$) and the local environment of the moir\'e superlattice at these specific points mimics the limiting configurations more and more as the twist angle decreases.
The lower limit of the distribution shifts to smaller values (\autoref{fig:structural-properties}a) but even at the lowest twist angle considered, the smallest values are still slightly larger than the interlayer distance for the limiting $R^X_h$ and $R^M_h$ structures where the interlayer distance attains its minimum (which corresponds to the lower limit of the x-range in \autoref{fig:structural-properties}a). 

One can also consider the distribution of the lateral displacements of the Mo atoms relative to the ideal moir\'e superlattice ($\delta d^{\text{Mo}}_{xy}$) that appear during relaxation (\autoref{fig:structural-properties}b-e).
The largest twist angle shown is 7.34$^{\circ}$ in which the maximal lateral displacement of the Mo atoms is \SI{0.025}{\angstrom}.
For the smallest twist angle (4.41$^{\circ}$), the maximum relaxation distance of the Mo atoms is significantly larger at \SI{0.065}{\angstrom}.
For the $4.41^{\circ}$ twist angle, the planar displacements forms three bands (\autoref{fig:structural-properties}e) corresponding to different distances of the displaced atom from the rotation center with a local AA stacking order.
The direction of the lateral displacements for the Mo atoms is shown in \autoref{fig:local-registry}b.
The regions around the high symmetry points are almost vertically aligned and exhibit very little relaxation.
The main relaxation occurs between high symmetry point $R_h^h$ and $R_h^X$ (and between $R_h^M$ and $R_h^h$).

\begin{figure}
    \centering
    \includegraphics[]{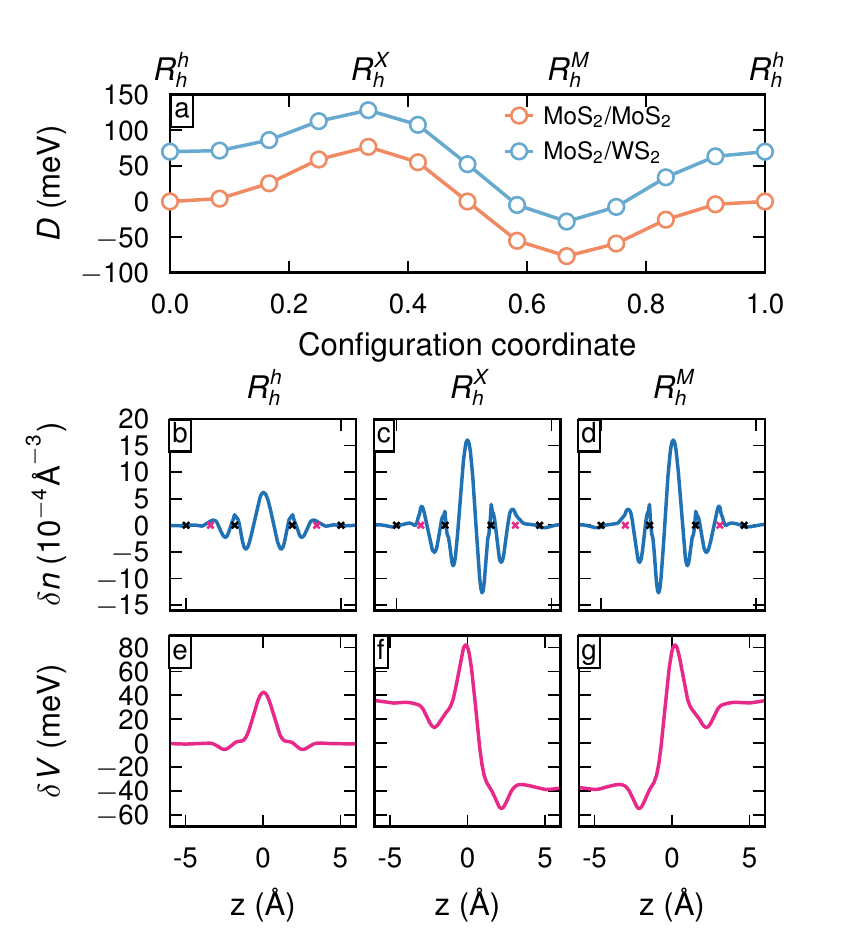}a
    \caption{
        (a) Schematic illustration of the vacuum level difference (\autoref{eq:D}) obtained by displacing the upper layer along the long diagonal of the primitive cell, keeping the interlayer distance fixed.
        (b-d) electron density ($\delta n$ of \autoref{eq:density-difference}) averaged in-plane for configurations $R_h^h$, $R_h^X$, and $R_h^M$ of \ce{MoS2}/\ce{MoS2}.
        (e-g) Corresponding potential $\delta V$ averaged in-plane.
    }
    \label{fig:density-potential}
\end{figure}

\subsection{Moir\'e potential}

Next we turn to the moir\'e potential.
In \autoref{fig:density-potential}a, a schematic illustration of the vacuum level difference (\autoref{eq:D}) as a function of displacement along the long diagonal of the unit cell is shown.
The minimum occurs at $R^X_h$ while the maximum occurs at $R^M_h$.
The origin of the spatially varying vacuum level difference is the stacking dependent density polarization (\autoref{fig:density-potential}b-d), which gives rise to asymmetric (with respect to the plane halfway between the monolayers) potentials (\autoref{fig:density-potential}e-g).
The potentials shown in (\autoref{fig:density-potential}e-g) would correspond to the \emph{interlayer} moir\'e potential at these specific sites within the moir\'e superlattice and the magnitude of this potential is around \SI{80}{\milli\electronvolt} for the \ce{MoS2}/\ce{MoS2} homobilayer.
Due to symmetry, this is also the full variation of the \emph{intralayer} moir\'e potential, and taken together it places the full variation of the \emph{interlayer} moir\'e potential at around \SI{160}{\milli\electronvolt} for this system. 

\begin{figure}
    \centering
    \includegraphics{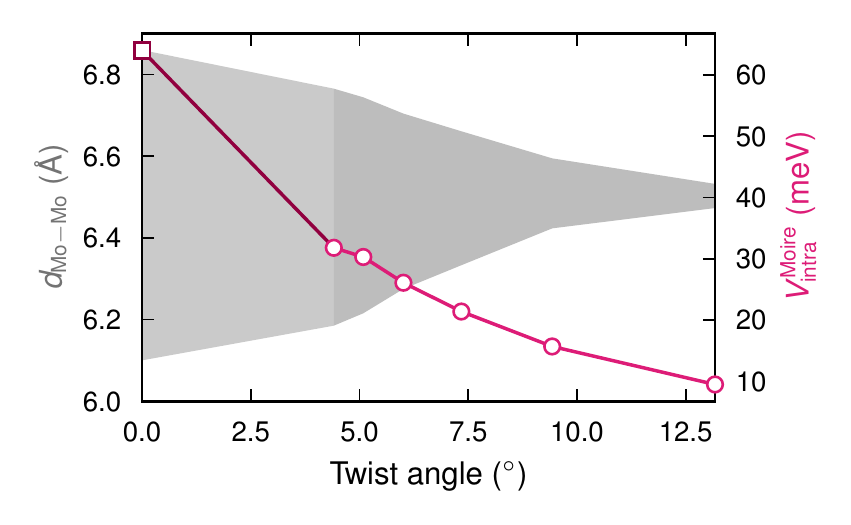}
    \caption{
        Maximal planar variation of the core potential for \ce{MoS2}/\ce{MoS2} moir\'e superlattices.
        $4.41^{\circ}$ is the smallest twist angle for which explicit computations have been performed.
        The data point (indicated by a square) at 0$^{\circ}$ is the zero degree limit computed as the Mo core potential difference at $R^h_h$ and the connecting line is a guide to the eye.
        The gray area indicates the range between the smallest and largest interlayer distance.
    }
    \label{fig:v-moire-intra}
\end{figure}

We have also studied the twist angle dependence of the \emph{intralayer} moir\'e potential by considering the maximal planar variation of the core potentials (\autoref{fig:v-moire-intra}).
We find that it decays rapidly with increasing twist angle and already at $\sim5^{\circ}$ it has decayed to half of the zero limit value.
Part of this decay can likely be explained by the increasing minimal interlayer distance with increasing twist angle and partly by the deviation from ideal stacking at $R_h^X$ and $R_h^M$.
This behavior of the moir\'e potential is consistent with explicit calculations of the \ce{MoS2}/\ce{MoTe2} system \cite{GenWanLin21}, with measured energy barriers of exciton diffusion in twisted \ce{MoSe2}/\ce{WSe2} \cite{LiLuCor21} and in the localization of excitons in \ce{MoS2}/\ce{WS2} \cite{GuaZhaXu20}.

The twist angle dependence of the moir\'e potential was computed without dipole corrections.
In order to validate this approach we performed a test with an out-of-plane lattice vector of \SI{100}{\angstrom} and compared it with the \SI{70}{\angstrom} results.
We found that the maximal planar variation of the Mo core potential for the 6.01$^{\circ}$ structure was \SI{26.1}{\milli\electronvolt} in both cases confirming that the spurious electric field does not influence the results.

\subsection{Hybridization and tunneling in \texorpdfstring{MoS$_2$/MoS$_2$}{MoS2/MoS2}}

\begin{figure}
    \centering
    \includegraphics{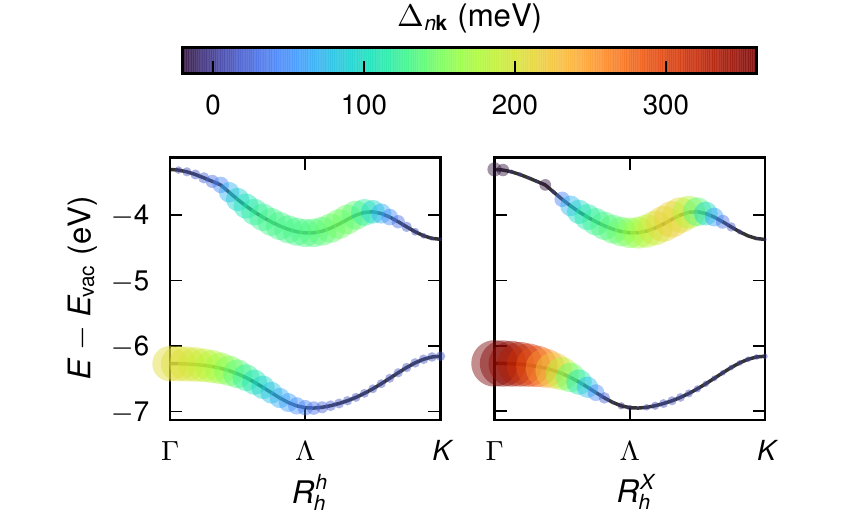}
    \caption{
        Average hybridization contribution according to Eq.~\eqref{eq:deltank} of the valence band and conduction band for the limiting configurations of \ce{MoS2}/\ce{MoS2} between the $\Gamma$ and ${K}$ points superimposed on the monolayer bands computed within the \gls{lda}.
    }
    \label{fig:deltank}
\end{figure}

The electrostatic moir\'e potential rigidly shifts the single particle electronic states according to Eq.~\eqref{eq:m}.
The bilayer states are, however, subject to interlayer hybridization as well, which for \ce{MoS2}/\ce{MoS2} homobilayer is quantified in \autoref{fig:deltank} along the path between the ${\Gamma}$ and ${K}$ points.
In all configurations, the hybridization contribution to the valence band shifts is small at the ${K}$ point and large at ${\Gamma}$.
For the conduction band, the hybridization contribution is largest at ${\Lambda}$ (halfway between ${\Gamma}$ and ${K}$).
At the $\Gamma$ point the hybridization is much stronger in $R_h^X$ and $R_h^M$ than for $R_h^h$.
At $K$, on the other hand, the valence band hybridization is slightly larger in $R_h^h$ (\SI{15}{\milli\electronvolt}) compared to $R^X_h$, and $R^M_h$ (\SI{5}{\milli\electronvolt}) despite exhibiting a much larger interlayer distance.
For the conduction band at $K$ in $R^X_h$, and $R^M_h$ the hybridization contribution is negative with a value of \SI{-2}{\milli\electronvolt}.

The tunneling rates defined in Eq.~\eqref{eq:t} largely follow the hybridization energy but become imaginary when the hybridization energy is negative, which implies that the solution becomes non-physical (\autoref{sect:discussion}).
This occurs at the $\Lambda$ point for the valence band and the $K$ point for the conduction band in $R_h^X$, and $R_h^M$.
The tunneling rate at $\Lambda$ for the conduction band is \SI{132}{\milli\electronvolt} in $R_h^h$ rising to \SI{184}{\milli\electronvolt} in $R^X_h$ and $R^M_h$.
The tunneling rate for the ${{K}}$ point is \SI{2}{\milli\electronvolt} (conduction band) and \SI{15}{\milli\electronvolt} (valence band) in $R_h^h$.
The latter value rises to \SI{19}{\milli\electronvolt} in $R^X_h$ and $R^M_h$ while the former becomes imaginary.
It was shown in Ref.~\citenum{TonYuZhu17} that the tunneling rate at ${K}$ vanished in $R^X_h$ and $R^M_h$ under the assumption that the orbital character of the monolayer band edges were completely composed of transition metal $d$ states.
We attribute the appearance of a finite tunneling rate for the valence band at $R^X_h$ and $R^M_h$ at ${K}$ to orbital mixing with other states.
Based on semi-local \gls{dft} calculations, it was found in Ref.~\citenum{HaaStrPan18} that there is a minor but non-zero contribution to the band edge states coming from the chalcogen species.

\begin{figure}
    \centering
    \includegraphics{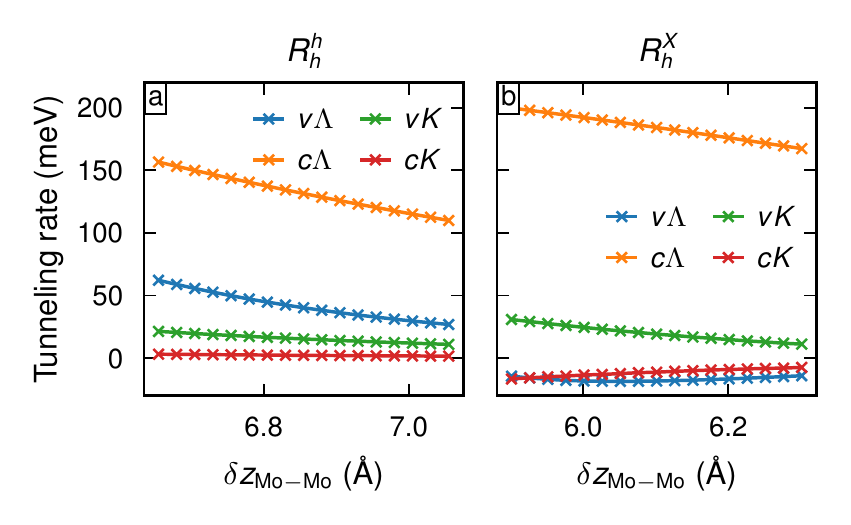}
    \caption{
        Tunneling rates as a function of interlayer distance for (a) $R_h^h$ and (b) $R^X_h$ of \ce{MoS2} homobilayer.
    }
    \label{fig:distance-T}
\end{figure}

The tunneling rates at ${\Lambda}$ and ${K}$ for $R_h^h$ and $R^X_h$ are to a good approximation linear in the interlayer distance (\autoref{fig:distance-T}).
In $R_h^h$, the interlayer dependence is stronger for both the states at ${\Lambda}$ compared with ${K}$, whereas the dependence is very weak for the conduction band states.
In $R^X_h$, the conduction band at ${\Lambda}$ and the valence band at ${K}$ exhibit stronger interlayer dependence than the other two states considered.

\begin{figure*}
    \centering
    \includegraphics[width=\linewidth]{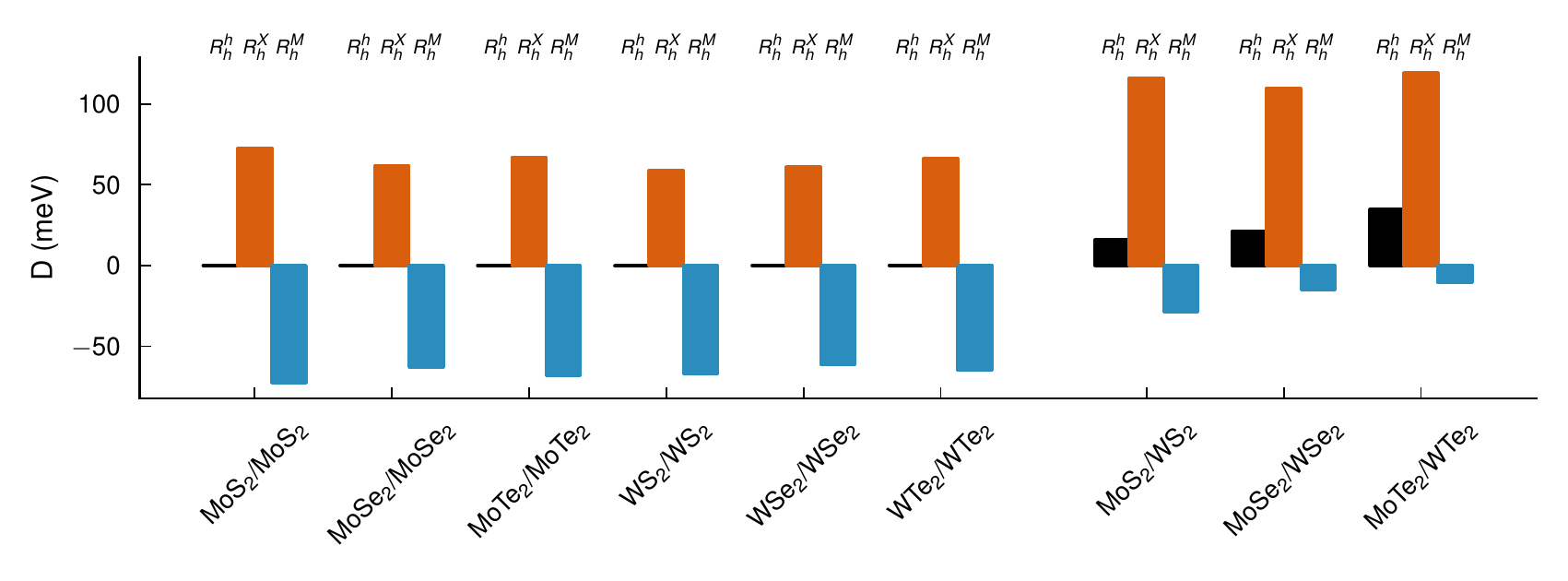}
    \caption{
        Maximal variation of the dipole field for homobilayers and heterobilayers at the limiting configurations according to Eq.~\eqref{eq:D}.
    }
    \label{fig:tmds-d}
\end{figure*} 

\subsection{Extension to other TMDs} \label{sec:other-tmds}

In \autoref{fig:tmds-d} the vacuum level difference $D$ as defined in Eq.~\eqref{eq:D} is shown for all Mo and W-based \glspl{tmd} including the heterobilayers.
For the homobilayers, the magnitude of the vacuum level difference is more or less uniform with minor variations between different \glspl{tmd} despite rather large differences in the interlayer distance.

For the heterostructures, in contrast to the homobilayers, there is a vacuum level splitting present in $R_h^h$ due to the presence of different transition metals.
However, the full variation of the interlayer moir\'e potential is still rather similar over both the homobilayers and heterobilayers with an average of \SI{131}{\milli\electronvolt} (standard deviation is \SI{8}{\milli\electronvolt}).
This is likely due to the fact that the relevant orbital character of all the \glspl{tmd} considered are similar.

\subsection{Excitons in TMD bilayers}

We have briefly investigated how the layer dependent polarization and hybridization influence the exciton spectrum in the different stackings originating from the AA stacked system. 
The present analysis has been performed by solving the Wannier equation in a basis of the constituent monolayers (see Eq.~\eqref{eq:hamiltonian}) and the resulting excitons contains both intra and inter-layer components \cite{OveBreLin19, MerMooSte19} (\autoref{fig:schematic-exciton}). 

\begin{figure}
    \centering
    \includegraphics[width=\linewidth]{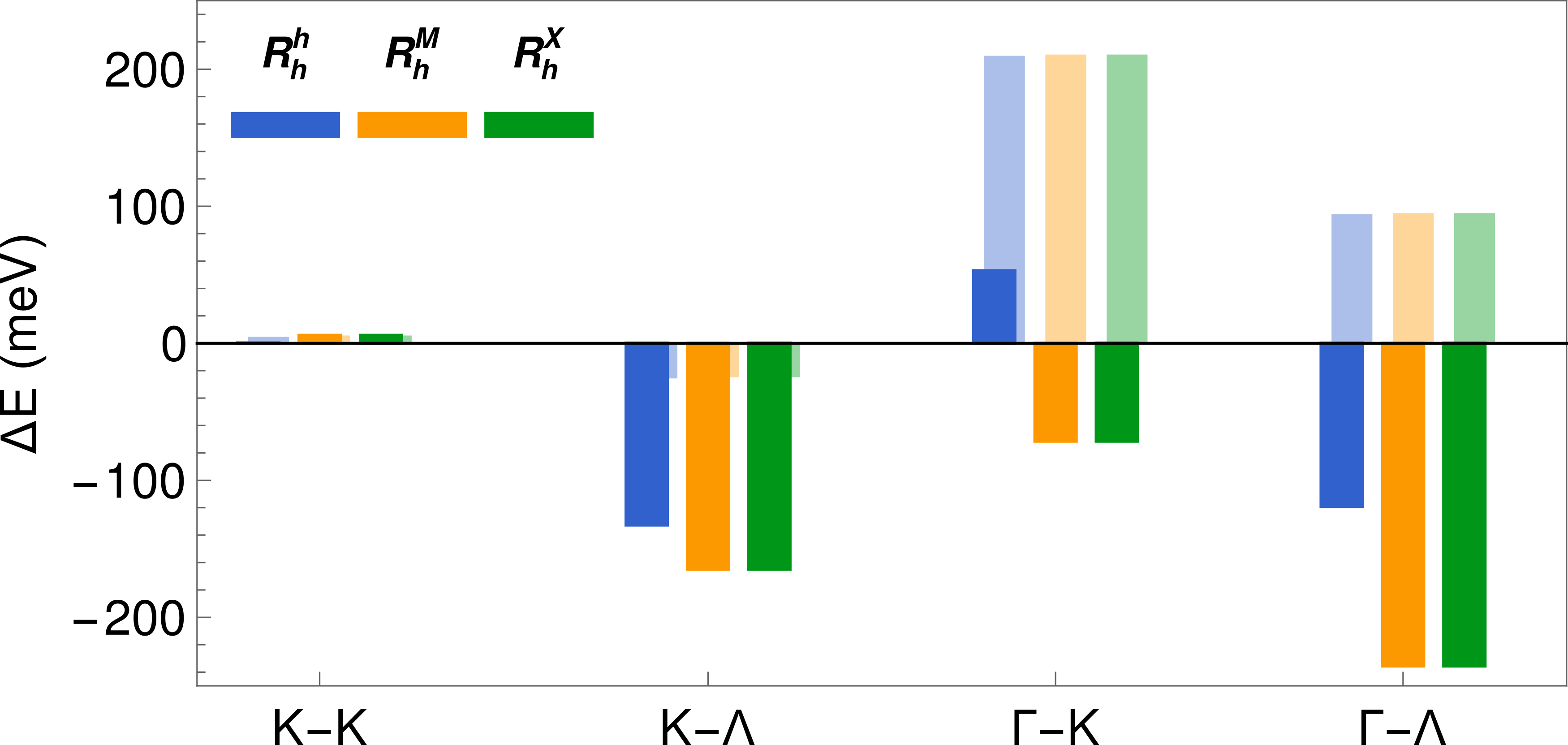}
    \caption{
        Energy of the lowest exciton peak relative to the lowest $K(\mathrm{hole})-K(\mathrm{electron})$ exciton for the WS$_2$ homobilayer.
        The shaded bars indicate the influence of the polarization shift and the solid bars indicate the influence of both the polarization shift and hybridization.
    }
    \label{fig:valley-homo}
\end{figure}

The exciton spectra were investigated for the two prototype cases $i)$ homobilayer WS$_2$, and $ii)$ the MoSe$_2$/WSe$_2$ heterobilayer. 
In \autoref{fig:valley-homo}, the energy of the lowest lying exciton peak relative to the $K-K$ exciton in the $R_{h}^h$ stacked structure is shown for the WS$_2$ homobilayer, and in \autoref{fig:valley-hetero}, the same data is shown for the MoSe$_2$/WSe$_2$ heterobilayer.
The transparent bars include the polarization shift but not the hybridization energy.
For the homobilayer, the lowest $K-K$ exciton is of intralayer nature and hence the stacking dependent polarization does not affect the position of the peak.
In particular, there is a very weak hybridization of the $K-K$ excitons. 
For the case of excitons that involve the $K-\Lambda$ valleys, the energy is around \SI{150}{\milli\electronvolt} lower than the $R_{h}^h$ $K-K$ exciton with the largest contribution stemming from the hybridization of the conduction band at the $\Lambda$ point.
To properly see the effect of hybridization, we consider the $\Gamma-K$ exciton in the $R_h^M$ stacking.
Without accounting for hybridization, the exciton energy is about \SI{200}{\milli\electronvolt} higher in energy in comparison with the $R_h^h$ $K-K$ exciton, since the \gls{vbm} at $\Gamma$ is lower in energy in comparison with at the $K$-point.
The \gls{vbm} at $\Gamma$ is pushed to higher energies by hybridization and the exciton energy becomes lower than the $R_h^h$ $K-K$ exciton.
The $\Gamma-\Lambda$ indirect exciton is the one mostly affected by the hybridization since it is strong in both these valleys.

In the heterobilayer however, where the lowest exciton exhibits interlayer character, there is a pronounced effect of the polarization field on the lowest $K-K$ exciton and the exciton energy difference between the different stacking orders corresponds to the variation of the dipole field.
The $K-\Lambda$ exciton is, without accounting for hybridization located around \SI{150}{\milli\electronvolt}-\SI{200}{\milli\electronvolt} above the $R_h^h$ exciton.
This energy drops significantly due to hybridization pushing the $\Lambda$ valley to lower energies.
In contrast to the case of the homobilayer where the indirect $\Gamma-K$ exciton is lower in energy with respect to the $R_h^h$ $K-K$ exciton, the $\Gamma-K$ exciton in the heterobilayer has a much larger energy in comparison with the $R_h^h$ $K-K$ exciton.

\section{Discussion}
\label{sect:discussion}

The main result of this study is that the moir\'e potential exhibits a strong twist angle dependence, as is evident from \autoref{fig:v-moire-intra}.
The origin of this potential is an asymmetric charge density displacement that is dependent on the twist angle.
The twist angle dependence of the potential can be traced back to the varying interlayer distance and the horizontal alignment. 
The latter was suggested to be the explanation for the twist angle dependence of the band gap in the twisted heterostructure \ce{MoS2}/\ce{MoTe2} \cite{GenWanLin21} and is likely to be the dominating contribution to the twist angle dependence of the moir\'e potential in \ce{MoS2}/\ce{MoS2} as well.

The implications of a twist angle dependent potential are manifold.
For example, it determines for which angles the electronic states at the band edges become localized.
Furthermore, the twist angle dependence has implications for the temporal localization of charge carriers, since the potential is effectively an energy barrier. 
One of the charge carriers will be subject to an energy barrier induced by the dipole potential. For example, the transition from the lowest hole potential configuration to $R_h^h$ raises the hole energy by $D/2$.
With increasing twist angle the energy barrier becomes smaller.
Therefore, it is expected that the temporal localization of charge carriers is strongest at very small twist angles.

Exciton migration may proceed via resonant energy transfer \cite{GuaZhaXu20} but in the case that electron and hole migrate separately the current results are consistent with measurements on exciton diffusion. 
The interlayer exciton diffusion in MoSe$_2$/WSe$_2$ was investigated in Ref.~\citenum{ChoHsuLu20} and it was found that the exciton diffusion was considerably larger at a 3.5$^{\circ}$ twist angle than a 1$^{\circ}$ twist angle. 

The numerical values of the dipole field amplitude $D$ are quite similar for all the \glspl{tmd} considered here (\autoref{fig:tmds-d}).
The similarity of $D$ despite vastly different interlayer distances is interesting in itself and suggests that the dipole field is primarily a geometrical property, which, together with orbital similarity across this class of materials, results in a similar magnitude.
In fact, the interlayer distance and the induced density difference ($\delta n$) are not independent quantities.
The density difference forms mainly between the layers, creating a bonding electron density that is slightly polarized, which in turn gives rise to an electric field that balances the bonding electron density and dispersion interaction. 

\begin{figure}
    \centering
    \includegraphics[width = \linewidth]{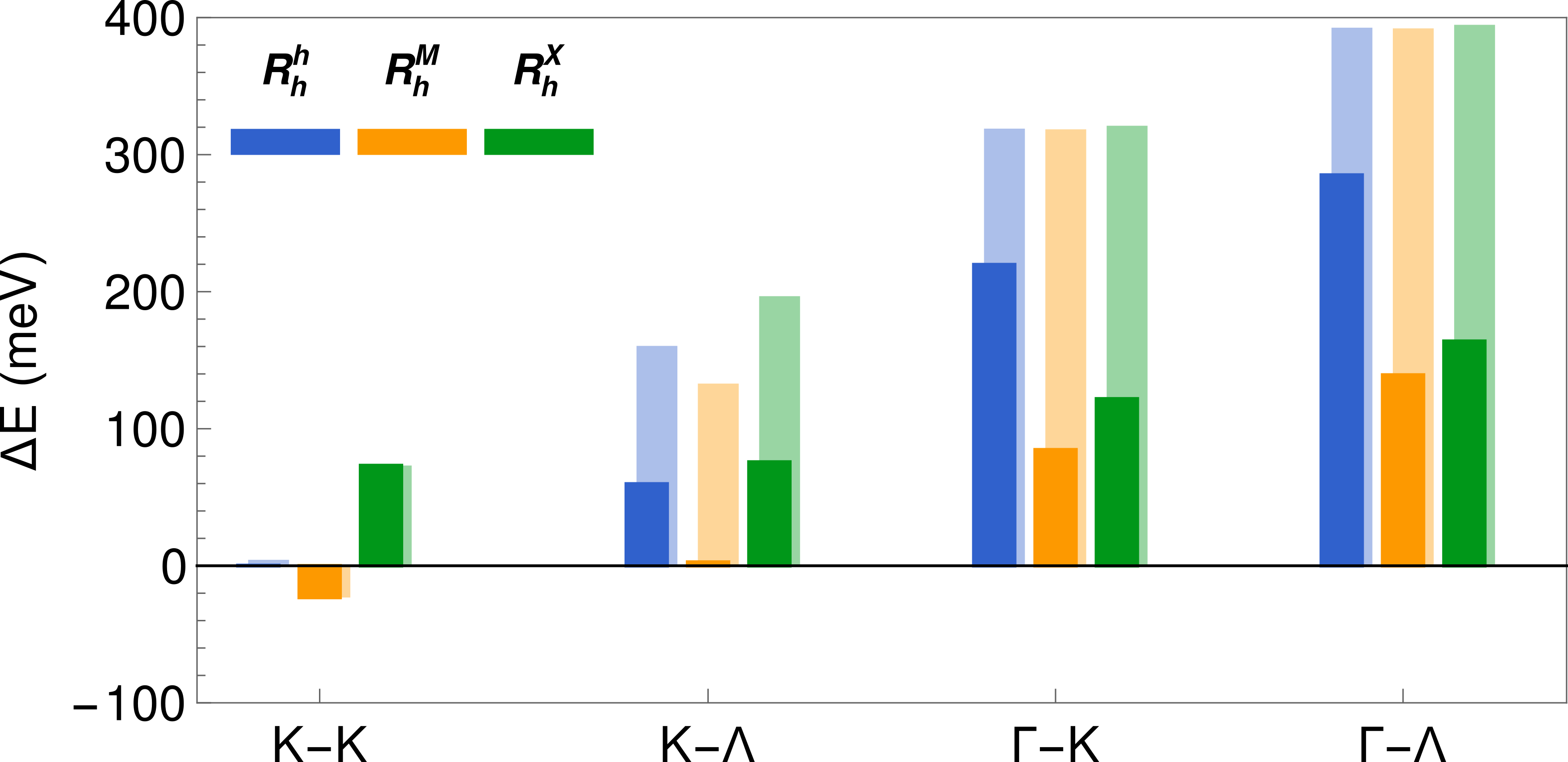}
    \caption{
        Energy of the lowest exciton peak relative to the lowest $K(\mathrm{hole})-K(\mathrm{electron})$ exciton for the \ce{MoSe2}/\ce{WSe2} heterobilayer.
        The shaded bars indicate the influence of the polarization shift and the solid bars indicate the the influence of both the polarization shift and hybridization.
    }
    \label{fig:valley-hetero}
\end{figure}

The moir\'e potential affects inter and intralayer excitons differently in the bilayer structures (\autoref{fig:valley-homo} and \autoref{fig:valley-hetero}) and the energy for the $K-K$ exciton with interlayer character exhibits an energy variation over the stacking orders $R_h^h$, $R^X_h$, and $R_h^M$ that closely resembles the dipole field magnitude.
The tunneling rates at the conduction band at $K$ and the valence band at $\Lambda$ are imaginary at $R^X_h$ and $R^M_h$ with similar values of around $10i-20i$~meV.
While the $\Lambda$ valence band is largely irrelevant due to its low energy, the conduction band at $K$ is rather important for the excitonic spectra.
The tunneling rates are, as constructed, off-diagonal components to a perturbation of the two-level system consisting of the equivalent band from different monolayers.
In this case, the splitting of the levels in the bilayer system is smaller than the difference in potential shift between the states.
The tunneling matrix element represents the interaction that causes band edge shifts that go beyond the splitting that is caused by potential alignment.
To describe this effect, here we employ a simple two-band model, see Eq.~\eqref{eq:t}.
The appearance of imaginary solutions can be attributed the breakdown of this approximation and signify that the bilayer energy shifts are influenced by coupling to multiple states. 

\section{Conclusions}

To summarize, we have shown that the moir\'e potential in \glspl{tmd} exhibits both an electrostatic component and a hybridization component, which are intertwined and have different relative strengths in different parts of the Brillouin zone.
The electrostatic component of the moir\'e potential is a varying dipole field, which has a strong twist angle dependence.
In some cases, the hybridization component can be interpreted as a tunneling rate but the interpretation is not generally applicable over the full Brillouin zone.

\begin{acknowledgments}
Funding from the Knut and Alice Wallenberg Foundation (2014.0226, 2019.0140) as well as the Swedish Research Council (2020-04935, 2018-06482) are gratefully acknowledged.
The Marburg group acknowledges support from Deutsche Forschungsgemeinschaft (DFG) via SFB 1083 (Project B9) and the European Unions Horizon 2020 research and innovation program under grant agreement No 881603 (Graphene Flagship).
The computations were enabled by resources provided by the Swedish National Infrastructure for Computing (SNIC) at C3SE and HPC2N partially funded by the Swedish Research Council through grant agreement no. 2018-05973.
\end{acknowledgments}

\end{document}